\documentclass[aps,rmp,twocolumn]{revtex4}

\usepackage{times}
\usepackage{graphicx}
\usepackage{amsmath}
\usepackage{color}
\usepackage{hyperref}

\newcommand{\EQ}[1]{Eq.~(\ref{eq:#1})}

\newcommand{\FIG}[1]{Fig.~\ref{fig:#1}}
\newcommand{\TAB}[1]{Tab.~\ref{tab:#1}}

\newcommand{\A}{\mathbf{A}}

\begin{document}
\title{Prediction, dynamics, and visualization of antigenic phenotypes of seasonal influenza viruses}

\author{Richard A.~Neher,$^{1}$ Trevor Bedford,$^{2}$ Rodney S.~Daniels,$^{3,4}$ Colin A.~Russell,$^{5}$ and Boris I.~Shraiman$^{6}$}
\affiliation{$^{1}$Max Planck Institute for Developmental Biology, 72076 T\"ubingen, Germany \\
$^{2}$Vaccine and Infectious Disease Division, Fred Hutchinson Cancer Research Center, Seattle, WA 98109, USA\\
$^{3}$The Francis Crick Institute, Worldwide Influenza Centre (WIC), Mill Hill Laboratory, The Ridgeway, London NW7 1AA, UK \\
$^{4}$formerly the WHO CC for Reference and Research on Influenza, MRC, National Institute for Medical Research, The Ridgeway, London NW7 1AA, UK\\
$^{5}$Department of Veterinary Medicine, University of Cambridge,
Cambridge CB3 0ES, UK \\
$^{6}$Kavli Institute for Theoretical Physics, University of California, Santa Barbara, CA 93106, USA
}
\date{\today}

\begin{abstract}
Human seasonal influenza viruses evolve rapidly, enabling the virus population to evade immunity and re-infect previously infected individuals. Antigenic properties are largely determined by the surface glycoprotein hemagglutinin (HA) and amino acid substitutions at exposed epitope sites in HA mediate loss of recognition by antibodies. Here, we show that antigenic differences measured through serological assay data are well described by a sum of antigenic changes along the path connecting viruses in a phylogenetic tree. This mapping onto the tree allows prediction of antigenicity from HA sequence data alone. The mapping can further be used to make predictions about the makeup of the future seasonal influenza virus population, and we compare predictions between models with serological and sequence data. To make timely model output readily available, we developed a web browser based application that visualizes antigenic data on a continuously updated phylogeny. 
\end{abstract}

\maketitle

Seasonal influenza viruses evade immunity in the human population through frequent amino acid substitutions in their hemagglutinin (HA) and neuraminidase (NA) surface glycoproteins \cite{hay_evolution_2001}. To maintain efficacy, vaccines against seasonal influenza viruses need to be updated frequently to match the antigenic properties of the circulating viruses. To facilitate informed vaccine strain selection, the genotypes and antigenic properties of circulating viruses are continuously monitored by the World Health Organization (WHO) Global Influenza Surveillance and Response System (GISRS), with a substantial portion of the virological characterizations being performed by the WHO influenza Collaborating Centers (WHO CCs) \cite{who_ssm_2014}.

\begin{figure}[hb!]
  \begin{center}
    \includegraphics[width=\columnwidth]{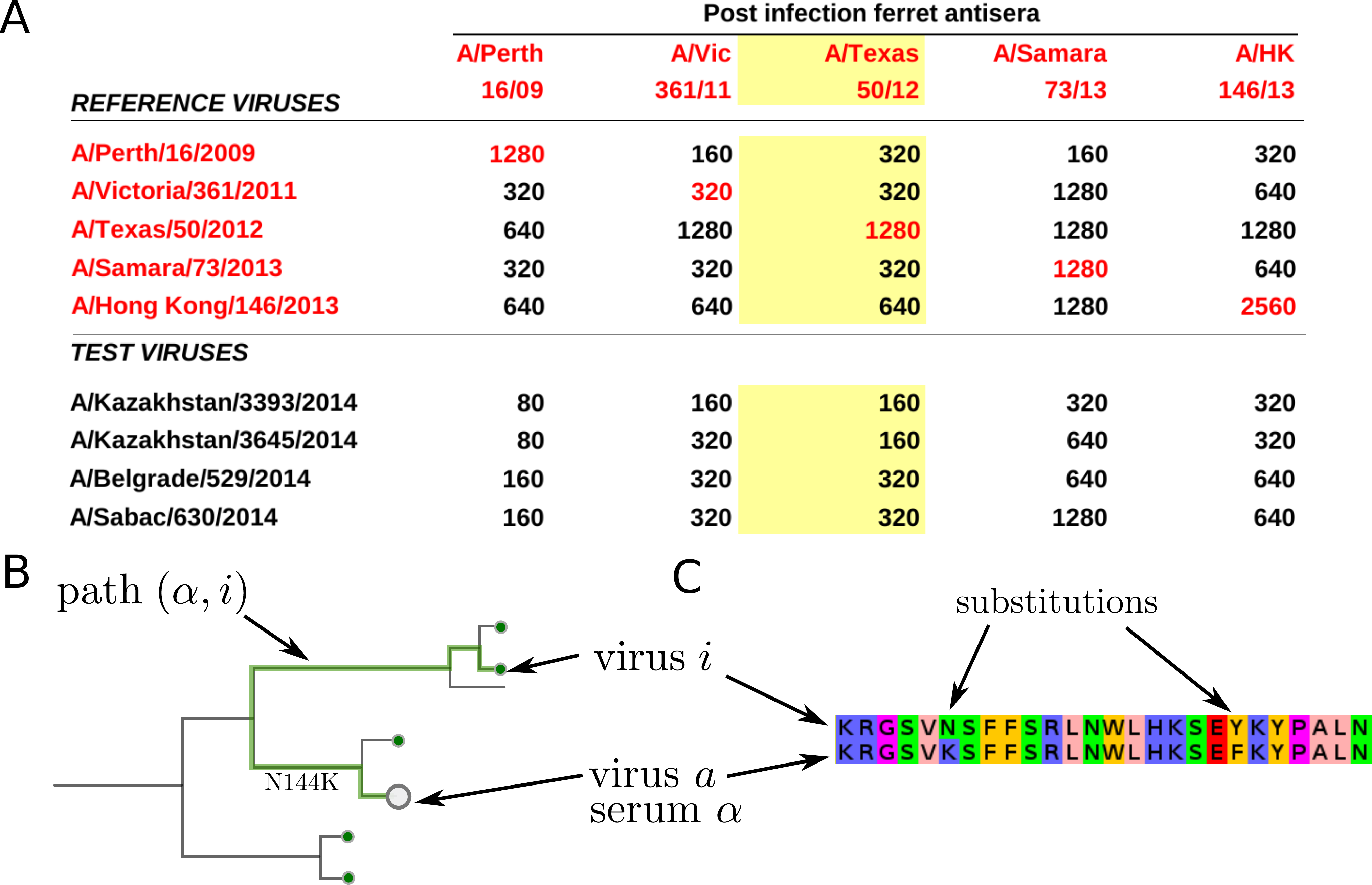}
  \end{center}
  \caption{{\bf Antigenic data and models for HI titers}. A: A typical table reporting HI titer data. Each number in the table is the maximum dilution at which the antiserum (column) inhibited hemagglutination of red blood cells by a virus (row). The red numbers on the diagonal indicate homologous titers. A typical HI assay consists of all reciprocal measurements of the available antisera and reference viruses, and a number of test viruses that are measured against all antisera, but for which no homologous antiserum exists. B: Each HI titer between antiserum $\alpha$ and virus $i$ can be associated with a path on the tree, indicated as a thick line. The tree model seeks to explain the titer distance as additive contributions of branches. C: In the substitution model, the sum over branches on the tree is replaced by a sum of contributions of amino acid substitutions.
  }
  \label{fig:illustration}
\end{figure}

Antigenic properties of influenza viruses are measured in hemagglutination inhibition (HI) assays \citep{hirst_quantitative_1942}, that record the minimal antiserum concentration (titer) necessary to prevent crosslinking of red blood cells by a standardized amount of virus based on hemagglutinating units (HAU). An antiserum is typically obtained from a single ferret infected with a particular reference virus. For a panel of test viruses, the HI titer is determined by a series of two-fold dilutions of each antiserum. An antiserum is typically potent against the homologous virus (the reference virus used to produce the antiserum), but higher concentrations (and hence lower titers) are frequently required to prevent hemagglutination by other (heterologous) viruses. HI titers typically decrease with increasing genetic distance from the homologous virus (\cite{hay_evolution_2001}).

Given multiple antisera raised against different reference viruses and a panel of test viruses, WHO CCs routinely measure the matrix of distances between a set of antisera and multiple test viruses, see \FIG{illustration}A. This distance matrix can be visualized in two dimensions via multidimensional scaling -- an approach termed antigenic cartography \cite{smith_mapping_2004}. While standard cartography does not use sequence information, sequences have been used as priors for distances in a Bayesian version of multidimensional scaling \cite{bedford_integrating_2014}. In order to infer contributions of individual amino acid substitutions to antigenic evolution, \citet{harvey_identifying_2014,sun_using_2013} have used models that predict HI titer differences by comparing sequences of reference and test viruses. 

Here, we show that antigenic distance relationships are accurately described by a model based on phylogenetic tree structure. We use the model to show that HI distances have a largely symmetric and tree-like structure. We show that large effect mutations account for about half of the total antigenic change and the effect of specific substitutions is strongly dependent on the genetic background in which they occur. We further investigate the ability of HI measurements to predict dominant clades in the next influenza season. To visualize antigenic properties on the phylogenetic tree, we have integrated the models of antigenic distances and the raw HI titer data into \href{http://HI.nextflu.org}{nextflu} -- an interactive real-time tracking tool for influenza virus evolution \cite{neher_nextflu:_2015}. 

This comprehensive summary of HA sequences from past and current influenza viruses linked to their antigenic properties has the potential to inform vaccine strain selections and facilitate efforts to predict successful influenza lineages \citep{luksza_predictive_2014,neher_predicting_2014,steinbruck_allele_2011,he_low-dimensional_2010,steinbruck_computational_2014}.

\section*{Results}
We use two related models that predict HI titers from sequences. The first  -- the \emph{tree model} -- expresses logarithmic HI titers by a sum of titer drops along branches in the phylogenetic tree that connect the test virus and the reference virus against which the antiserum was raised, see \FIG{illustration}B. The second -- \emph{the substitution model} -- explains HI titers as a sum of titer drops associated with amino acid substitutions between reference and test virus, see \FIG{illustration}C. Not all branches and not all substitutions are expected to result in changes to titer. These two models are similar but complement each other in a few aspects that we discuss further below. In addition to titer drops associated with branches or substitutions, individual antisera and viruses differ in their activity in HI assays. In the model, we account for this variability through `antiserum potency', which raises or lowers the expected titer of all HI measurements against an antiserum, and `virus avidity', which raises or lowers the expected titer of all HI measurements against a virus.

Specifically, in the tree model the antigenic distance $\Delta_{i\alpha}$ between virus $i$ and antiserum $\alpha$ is modelled as
\begin{equation}
\label{eq:model}
\Delta_{i\alpha} = p_\alpha + c_i + \sum_{b\in (i\ldots\alpha)} d_{b}
\end{equation}
where $d_b$, $p_\alpha$, $c_i$ denote the titer drop on a branch $b$ in the path $(i\ldots\alpha)$, the potency of antiserum $\alpha$, and the avidity of virus $i$, respectively. In the substitution model, the sum over branches is replaced by a sum over amino acid substitutions between reference and test virus. The parameters of our model are determined by minimizing a cost function that encourages a sparse model, i.e., a model that explains titers with a small number of large parameters and many parameters set to zero. This is achieved by $\ell_1$-norm regularization and non-negativity constraints, corresponding to the assumption that antigenic effects are positive and exponentially distributed. In contrast, the parameters absorbing variation among virus/antisera interactions are regularized by their $\ell_2$-norm, corresponding to a Gaussian prior on these parameters. For a detailed description of the models and inference procedures, see Material and Methods.

The decomposition of the HI data into antiserum potency, avidity of a virus, and a sparse genetic component (branches or substitutions) has the effect of smoothing the HI distance relationships and virus and antiserum specific variation is absorbed into the corresponding effects. The substitution or branch effects pick up titer drops associated with a larger number of antiserum-virus pairs.

\begin{figure}[tb]
  \begin{center}
    \includegraphics[width=0.48\columnwidth]{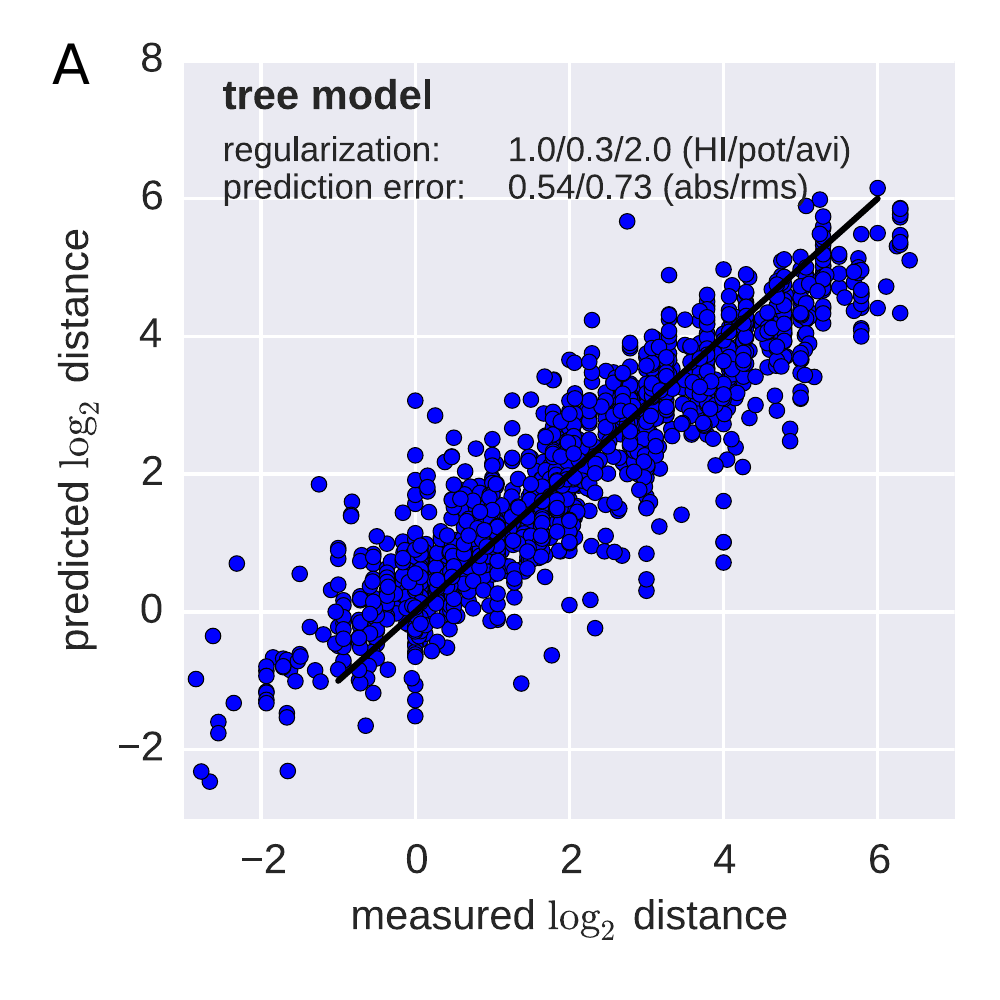}
    \includegraphics[width=0.48\columnwidth]{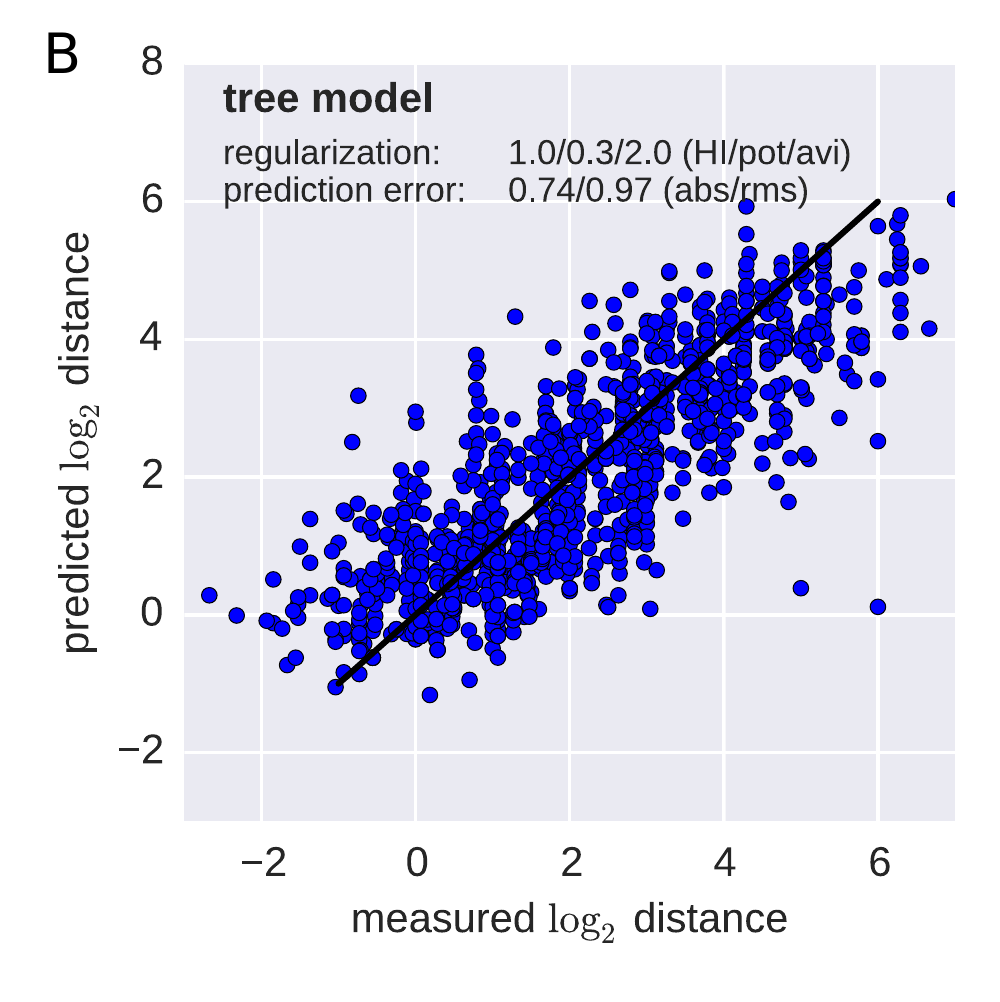}
  \end{center}
  \caption{{\bf HI titers are accurately predicted by the tree model}. The figure scatters predicted (y-axis) against a test set of measurements not used for training of the model (x-axis). This test set either consists of A) a random sample of 10\% of all measurements, or B) all measurements for 10\% of all viruses. In the latter, no avidity of a virus can be estimated for viruses in the test set since these viruses are completely absent from the training data. Hence prediction accuracy is lower but still comparable to the measurement accuracy.}
  \label{fig:titer_prediction}
\end{figure}

\begin{table}[tb]
  \caption{Prediction accuracy of the tree and substitution model. The terms \emph{titer} and \emph{virus} refer to predictions of individual titer measurements or all measurements for viruses completely absent from the training data, respectively. }
  \label{tab:prediction_accuracy}
  \centering

  \begin{tabular}{|l|rr|rr|}
  \hline
  \hline
  \textbf{model:} & \multicolumn{2}{c|}{tree}& \multicolumn{2}{c|}{substitution} \\
  \hline
  \textbf{test set:}& titer & virus& titer & virus \\
  \hline\hline
    A(H3N2) (12y)  & 0.54 & 0.72 & 0.53 & 0.71\\\hline
    A(H1N1pdm09) (7y)& 0.46 & 0.77 & 0.5 & 0.72\\\hline
    B/Yam (12y) & 0.68 & 0.86 & 0.65 & 0.88\\\hline
    B/Vic (12y) & 0.74 & 0.86 & 0.76 & 0.84\\\hline
  \hline
  \end{tabular}
\end{table}

\subsection*{The tree and substitution models accurately predict HI titers}
To evaluate the performance of the models we trained them on 90\% of the data and used the remaining measurements to validate the models as in \citep{bedford_integrating_2014}. We found that the models were able to predict titers of antiserum-virus combinations to an accuracy of approximately $0.5 \log_2$ titer levels for H3N2 with somewhat lower accuracy for the influenza B lineages (\TAB{prediction_accuracy}, \FIG{titer_prediction}A). Very little consistent antigenic evolution is observed in A(H1N1pdm09).

To quantify the prediction accuracy for viruses for which no antigenic data exists, we selected 10\% of the viruses and excluded all measurements involving these viruses from the training data. Both models predicted titers for viruses not part of the training set to an accuracy of approximately $0.75 \log_2$ titer levels (\TAB{prediction_accuracy}, \FIG{titer_prediction}B). Having completely excluded a virus from the training data implies that no avidity of a virus can be estimated. The increased prediction error is therefore largely due to virus-to-virus variability that is not captured by the HA phylogeny. 

In order to infer the genetic component of an HI titer, the relevant branches in the tree or the substitutions that separate test and reference virus have to be constrained by measurements in the training data set. For a completely novel clade in the tree, the model would predict HI titers equal to that of the base of the clade for all subtending viruses. Similarly, accurate inferences by the substitution model require the effects of the relevant substitutions to be constrained by training data.

Using the tree and substitution models, we can predict HI titers for every combination of antiserum and virus in a phylogenetic tree (with prediction confidence varying by quality and amount of antigenic data), essentially providing a smoothed interpolation of antigenic evolution on the tree. Note that the model correctly predicts titers in excess of homologous titers (negative values in \FIG{titer_prediction}A). These higher titers often coincide with large negative virus avidities, which explains the absence of titers predicted to be strongly negative in \FIG{titer_prediction}B, where no virus avidities are available.

\begin{figure}[tb]
  \begin{center}
    \includegraphics[width=\columnwidth]{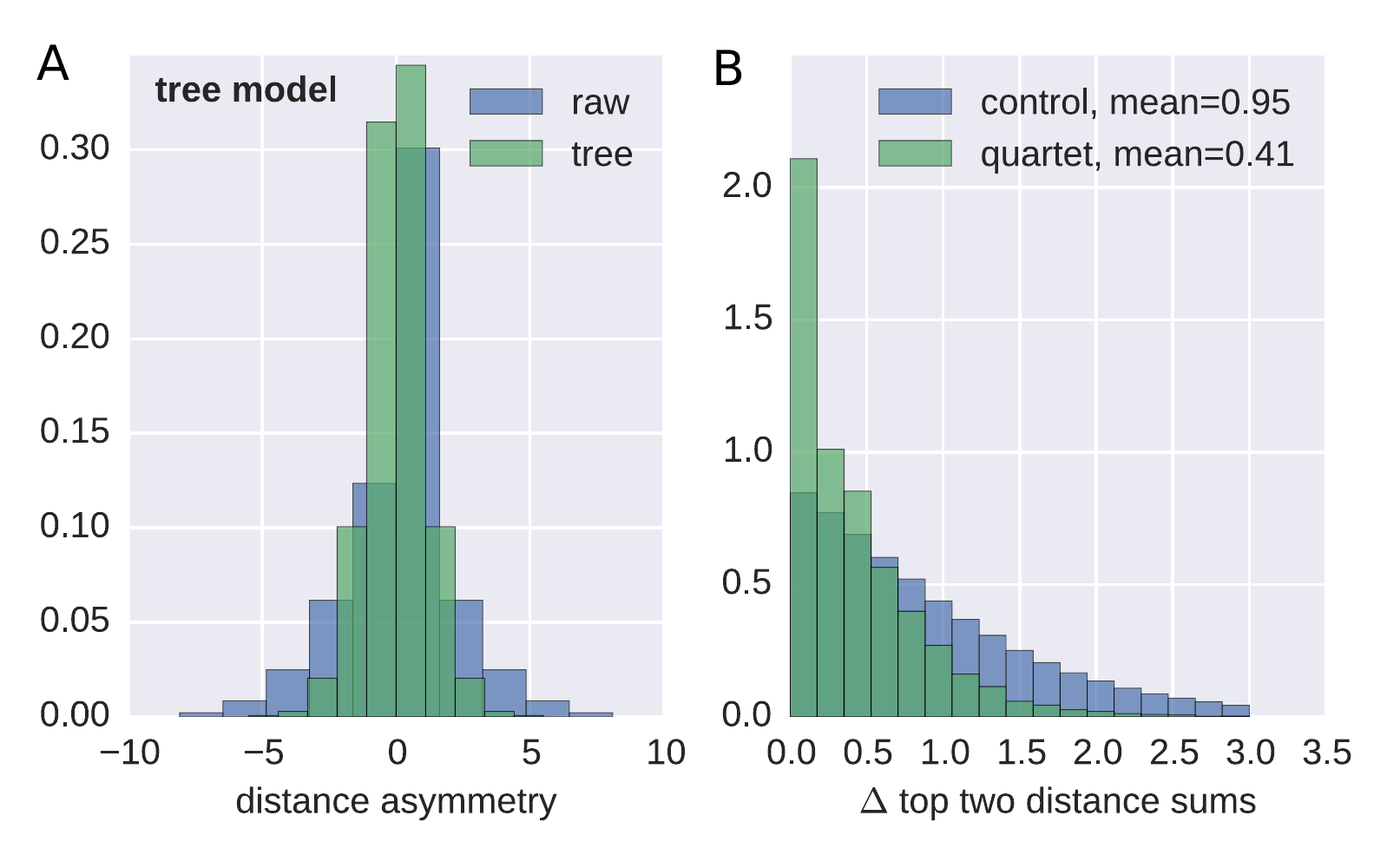}
  \end{center}
  \caption{{\bf HI titers approximate a distance on the tree.} A) Reciprocal HI titer measurements often differ by several $\log_2$ units. After subtracting the inferred virus and antiserum effects the remaining HI titer distance is almost symmetric with deviations again on the order of the measurement accuracy. B) On a tree, the two largest sums of distances within antisera / virus quartets are equal (see text). The figure shows the distribution of the absolute value of the difference of the top two distance sums for quartets and for three random distance sums. Those from quartets have a much smaller deviation.
  }
  \label{fig:symmetry}
\end{figure}

\subsection*{HI titers approximate a symmetric tree distance}
To further evaluate the approach of partitioning a titer into a antiserum, a virus, and a tree component, we considered all reciprocal titer measurements, i.e., pairs of viruses  $a,b$ against which antisera  $\alpha$, $\beta$ have been raised and that have been measured against each other. Subtracting the virus avidities and antiserum potency contributions from titers, the remainders $H_{a,\beta} - c_a - p_\beta$ and $H_{b,\alpha} -c_b -p_\alpha$ should reduce the titers to the symmetric tree component. \FIG{symmetry}A compares the distribution of $H_{a,\beta} - c_a -p_\beta - H_{b,\alpha} + c_b + p_\alpha$ with the uncorrected difference between the reciprocal titers $H_{a,\beta} - H_{b,\alpha}$. While raw reciprocal titer measurements often differ by several $\log_2$ titer units, the corrected tree component was symmetric to within one unit, as expected.

The degree to which titer distances have tree-like properties can be tested using the following quartet rule: take any four leaves $a,b,c,d$ and construct the sum of the distances $d_{ab}+d_{cd}$, $d_{ac}+d_{bd}$ and $d_{ad}+d_{bc}$. If the distances are given by a tree, the two largest of the three sums of distances will be equal. To test this quartet rule we determined maximal `cliques' of reference antisera, the activities of which had been measured against all viruses used to generate the antisera. Out of these cliques, we selected at random four antisera, and compared the two largest distance sums. As a control, we constructed three sums of two distances by three times randomly drawing four antisera. The difference between the largest and second largest of these distance sums has a mean of 0.41 $\log_2$ titer levels, while the control has a mean of 0.95 $\log_2$ titer levels (\FIG{symmetry}B). The similarity of the largest and second largest sums supports a tree-like structure underlying observed HI distances.

\subsection*{Amino acid substitutions associated with titer drops}
The majority of antigenic evolution tends to occur at a subset of sites \citep{shih_simultaneous_2007,munoz_epitope_2005} with seven positions near the HA receptor binding site (Koel 7) playing a particularly prominent role \citep{koel_substitutions_2013}. However, the number of substitutions at these sites is a poor predictor of antigenic distance with $r^2$ values of about 0.25  (supplementary \FIG{distance_correlations}), possibly because the effect of a substitutions depends on the genetic background (HA sequence) in which it occurs. 

Within the substitution model, we can explicitly associate antigenic changes with amino acid substitutions. \TAB{aminoacids} shows the top 50 contributions for A(H3N2) for sequences estimated from sets of sequences and the associated HI titers in five overlapping 10 year intervals. Most of the largest contributions coincide with substitutions at Koel 7 sites. When a substitution is present in the same context in overlapping time intervals, the estimated effects tend to be similar (K189N, K135E, K158R, K140E, Y159F) -- provided there is enough data to constrain the model. 

By and large, the effects we infer are compatible with those associated with cluster transitions determined in \citep{koel_substitutions_2013}. For example, the transition from the Sichuan/87 (SI87) cluster to the Beijing/89 (BE89) cluster involved the substitution N145K, for which we infer an effect of 1.95 antigenic units (each unit is equal to a two-fold decrease in HI titer). Together with minor effects associated with N193S, I186S, and G135N, this comes close to the map distance 3.9 units. For the transition from SI87 to Beijing/92 (BE92), the substitution model estimates a distance of 4.5 that is associated with many almost simultaneous substitutions -- hence the exact assignment remains difficult to pin down. The map distance corresponding to this cluster transition is 7.8 units, while the typical titer drop is approximately 5 units and hence closer to the estimate of the tree or substitution model. The transition from BE92 to Wuhan/95 (WU95) is gradual with several substitutions of intermediate magnitude. From WU95 to Sydney/97 (SY97), we estimate a distance of approximately 3 units of which 2 units are attributed to the set of substitutions: K62E/V144I/K156Q/E158K/V196A/N276K. These substitutions account for 3.6 units in cartography. Lastly, the transition from SY97 to Fujian/02 (FU02) is attributed to Q156H (1.05 units as in \citep{koel_substitutions_2013}) and N121T. Most of the distances we infer between clusters are smaller than those estimated in cartography, which is compatible with our finding of overall reduced antigenic drift when modeling antigenic evolution in many dimensions (see below). 

The inferred effects of substitutions suggest substantial dependence on genetic background and the specific amino acid change. For HA1 position 145, the back and forth between asparagine (N) and lysine (K) is associated with different effects in 1989--1992, and 2004. Interestingly, the inferred effect of the forward substitutions (N145K in 89--92 and K145N in 04) is large in these instances, while the inferred effect for the backward direction is small. Some substitutions at Koel 7 positions have very small effects, as for example E158D and S189N which involve biochemically similar amino acids. Other substitutions at these positions have large effects, but don't spread. K158R, for example, shows up repeatedly and is associated with a 2 unit titer drop without ever reaching high frequencies. A full table of all inferred effects for substitutions at particular positions in overlapping 10 year intervals is given as supplementary Tab.~2. However, the interpretation of effects is sometimes difficult due to colinearities between substitutions.

\begin{figure}[tb]
  \centering
  \includegraphics[width = 0.9\columnwidth]{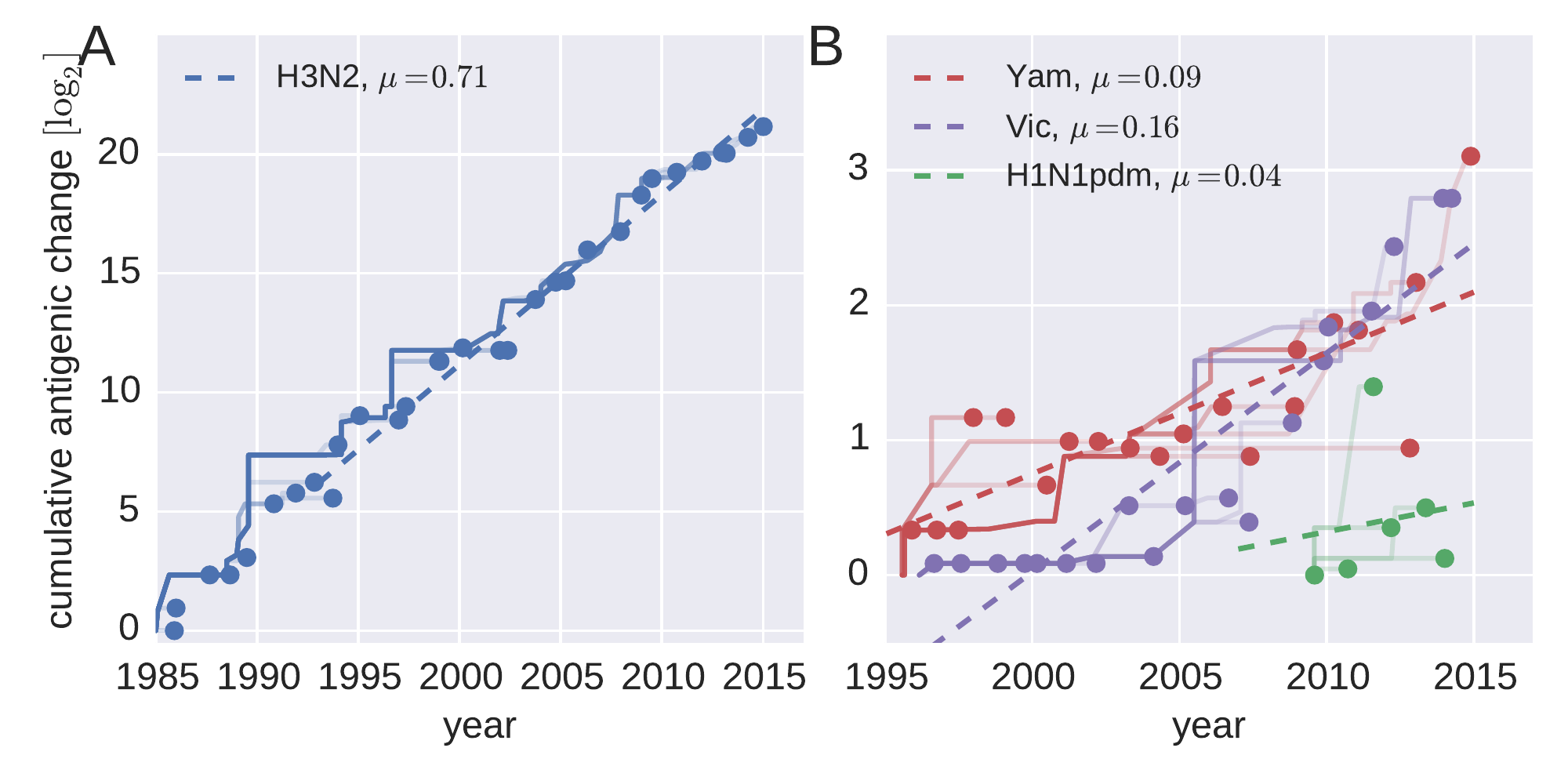}
  \caption{{\bf Cumulative antigenic change from the root to the leaves of the tree.} A)  shows H3N2 antigenic evolution over the past 30 years. B) shows the corresponding traces for the two influenza B lineages (past 20 years) and H1N1pdm09 (past 7 years).}
  \label{fig:cumulative}
\end{figure}
\begin{figure*}[tb]
\raggedright
  \includegraphics[width=0.9\linewidth]{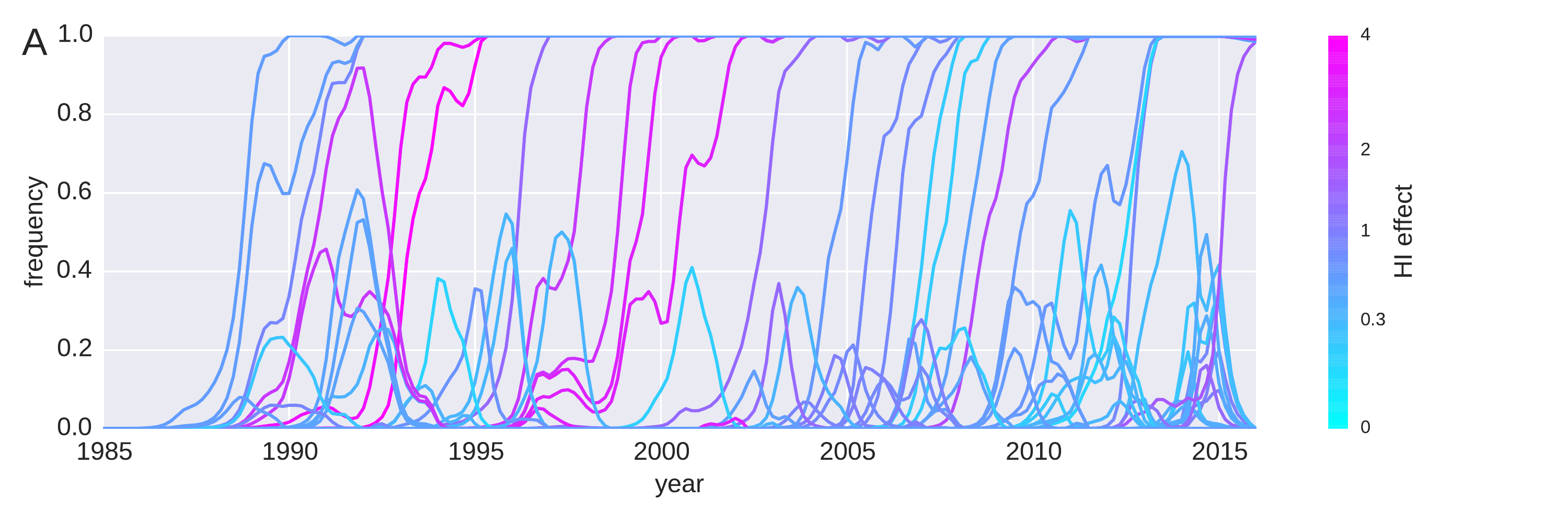}
  \includegraphics[width=0.75\linewidth]{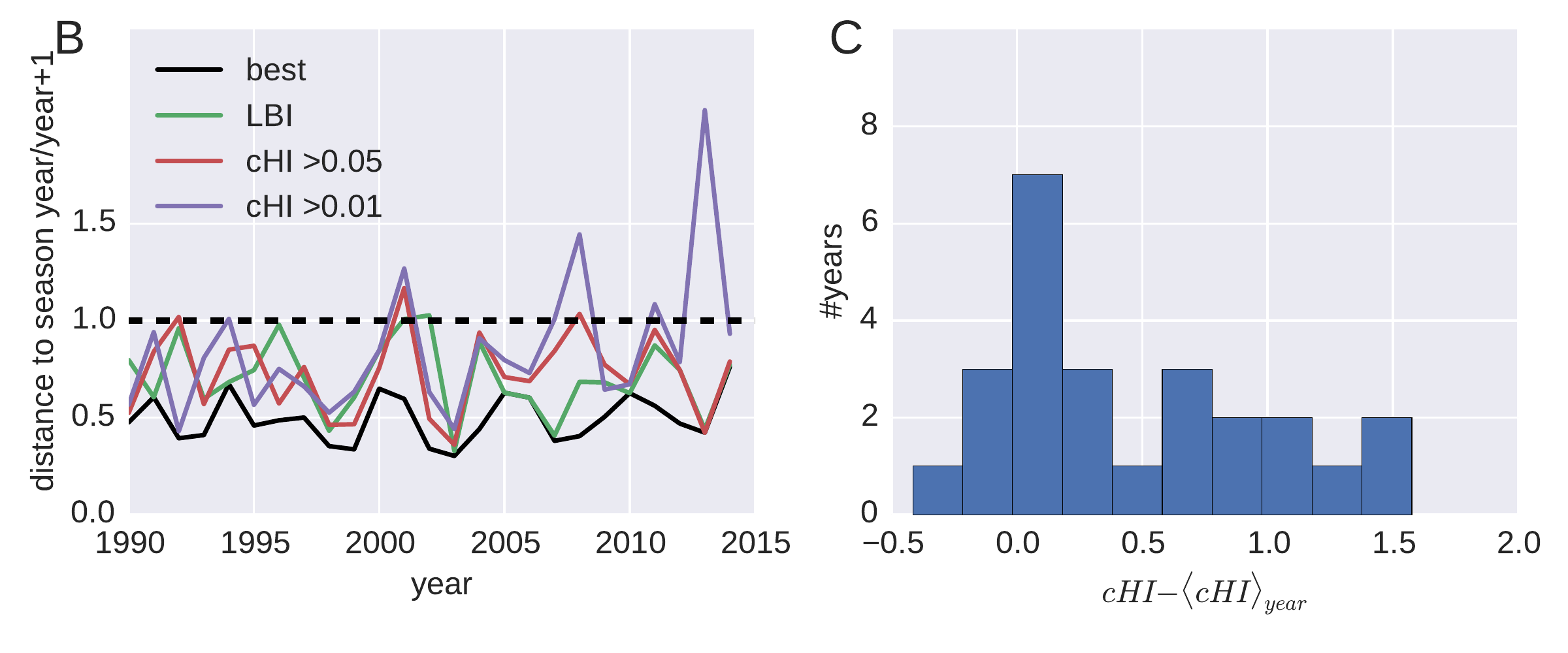}
  \caption{{\bf Antigenic evolution and the success of clades.} A) shows trajectories of clades colored by inferred cumulative antigenic evolution in the past 6 months on branches ancestral to the clade. The majority of the clades with very large HI effects fix, but several branches with substantial antigenic evolution go extinct. B) shows genetic distances to the next season of nodes with maximal LBI or cHI (distance is normalized such that the average clade has distance 1). The black line indicates the distances of the best possible pick. Predictions based on maximal LBI and cHI are correlated, but maximal cHI is sensitive to false positives. C) shows the histogram of centered cHI of the clade closest to the next season for each year, suggesting that successful clades tend to be antigenically advanced.}
  \label{fig:spread}
\end{figure*}

\subsection*{Cumulative antigenic evolution}
By summing all contributions to antigenic change on branches on the path between a virus and the root of the tree, we can estimate the total past antigenic change. This cumulative antigenic change is roughly comparable to `dimension 1' in antigenic cartography (more precisely analogous to the length of a path on the map corresponding to the trunk of the tree).
H3N2 viruses advance by approximately 0.7 $\log_2$ titer units per year (\FIG{cumulative}A), while influenza B virus lineages advance 0.16 (Vic) and 0.09 (Yam) units per year (\FIG{cumulative}B). Although consistent in relative rates across lineages, these absolute rates tend to be slightly lower than estimated previously using cartography \citep{bedford_integrating_2014}, possibly because forcing the antigenic distance matrix into two dimension causes distortions resulting in a slight exaggeration of the distances. Very little consistent antigenic evolution is observed in H1N1pdm09 (\FIG{cumulative}B).

We estimate that in A(H3N2) approximately half the cumulative antigenic evolution is due to a large number of substitutions with effects smaller than one unit, while 20\% is accounted for by a few substitutions with effects above two units (\FIG{cumulative_effects}). However, some of largest effects are associated with clusters of colinear amino acid changes and their individual effects cannot be resolved.

\subsection*{Limitations of the models}
The tree and substitution model perform similarly in terms of accuracy as summarized in \TAB{prediction_accuracy}, which is expected since branches of the tree are associated with substitutions and vice versa. In particular circumstances, however, one model is more accurate than the other and the two models complement each other in a number of situations.

The tree model assumes that titers are additive along the tree. While this is in general a reasonable assumption, it is violated when the same amino acid position is mutated multiple times in different parts of the tree. In a recent example from H3N2, there are independent substitutions at HA1 site 159 with F159Y resulting in clade 3c2.a viruses and F159S resulting in clade 3c3.a viruses. The distance between 3c2.a and 3c3.a viruses is not necessarily a sum of the effects associated with the branches on which these two substitutions occur. In such cases, the substitution model tends to be more accurate as it has the additional freedom to introduce the Y159S substitution (and the reverse) that directly compares 3c2.a and 3c3.a viruses.

Similarly, the substitution model fails when the same substitution (as opposed to different substitutions at the same site) occurred in different places on the tree in different genetic backgrounds. For sequence ensembles covering long time periods, a substitution might have happened at different times in different genetic backgrounds. The model will fit a single effect, even though the effect of the substitution might be background dependent. Furthermore, the substitution model tends to be inaccurate when predicting titers for test viruses that predate the reference virus. Such `back-in-time' measurements are underrepresented in the data, and while the forward substitution might have a large effect assigned, the few back-in-time measurements do not provide enough support to include the reverse substitutions into the model. The tree model does not suffer from this problem, as effects are assumed to be symmetric.

By and large the model accurately predicts measured HI titers (\FIG{titer_prediction}) and deviations affect only isolated clades, typically when very few measurements are available to constrain the model. The visualization described below allows a direct side by side comparison of the two models and the measurements, which makes it easy to identify such isolated inaccuracies.

\subsection*{Antigenic change and the success of clades}
Antigenic changes result in viruses able to reinfect individuals with immunity to previously circulating viruses.
Intuitively, large antigenic changes should therefore be positively selected for and rapidly spread through the virus population, unless there is a fitness cost for the virus. We investigated the relationship between the amount of antigenic change and success of clades in the phylogenetic tree. \FIG{spread} shows frequency trajectories of clades reaching at least 10\% at one time as estimated by nextflu, while color indicates the magnitude of the antigenic change that accumulated along the ancestral lineage over the past six months. Although large antigenic changes appear to fix more often than small antigenic changes, there are also several clades that evolved antigenically but failed to spread. 

For each season, we determined the clade with highest Local Branching Index (LBI, a predictor of clade success \citep{neher_predicting_2014}) and the clade with the largest antigenic advancement (cHI) relative to all other viruses in a season (restricted to clades that account for at least 5\% of available sequences for the given season). 
For each of these clades, \FIG{spread}B shows genetic distance to the virus population of the following season.  Antigenic advance as measured by cHI is predictive of which lineage would dominate the following season: the distance is significantly below the population average. Predictions by cHI and LBI are comparable in quality and are correlated. Yet, clades with maximal cHI are sometimes far from the future population suggesting that a predictor based on antigenic phenotype alone readily generates false positives: the problem becomes worse when smaller clades are included (threshold lowered to $>1$\% of all sequences). Nevertheless, successful clades tend to be antigenically advanced. \FIG{spread}C shows the histogram of centered cHI for clades closest to the next season. These clades tend to have larger cHI than co-circulating viruses. These results suggest that cHI correlates with viral fitness but is not the only factor determining clade success. Supplementary \FIG{best_HI_vs_HI_of_best} compares the cHI of the most antigenically advanced clade to that of the successful clade in each year.

\begin{figure}[tb]
\raggedright
  \includegraphics[width=0.99\linewidth]{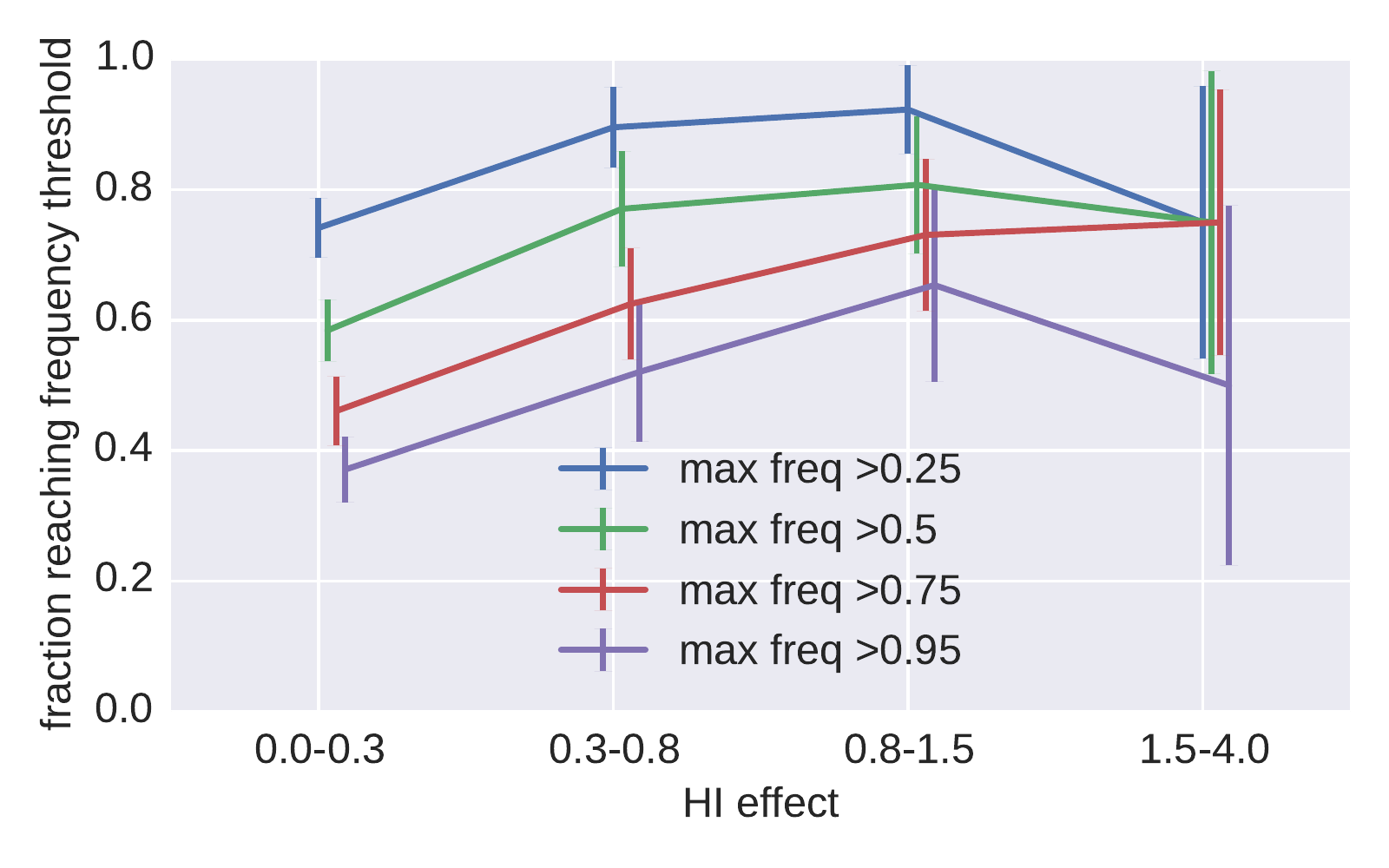}
  \caption{{\bf Substitutions with large antigenic effect fix preferentially}. The figure shows the fraction of all substitutions that reach the indicated frequency for different magnitudes of the inferred antigenic effect of this substitution. The plot combines data from substitution models fitted to 5 overlapping 10 year intervals from 1985 to 2015 and contains all substitutions that reach at least 10\% population frequency. There are few substitutions in the highest HI category -- error bars show the standard deviation over boot strap replicates of substitutions.}
  \label{fig:fraction_successful}
\end{figure}

As an alternative to clades in the tree, \FIG{fraction_successful} shows the probability of a substitution reaching a maximal frequency as a function of its inferred antigenic effect. This fixation probability increases with antigenic effect, but even substitutions with very large effect can fail to spread -- an effect which limits predictive power. 

\begin{figure}[tb]
  \centering
  \includegraphics[width=0.99\linewidth]{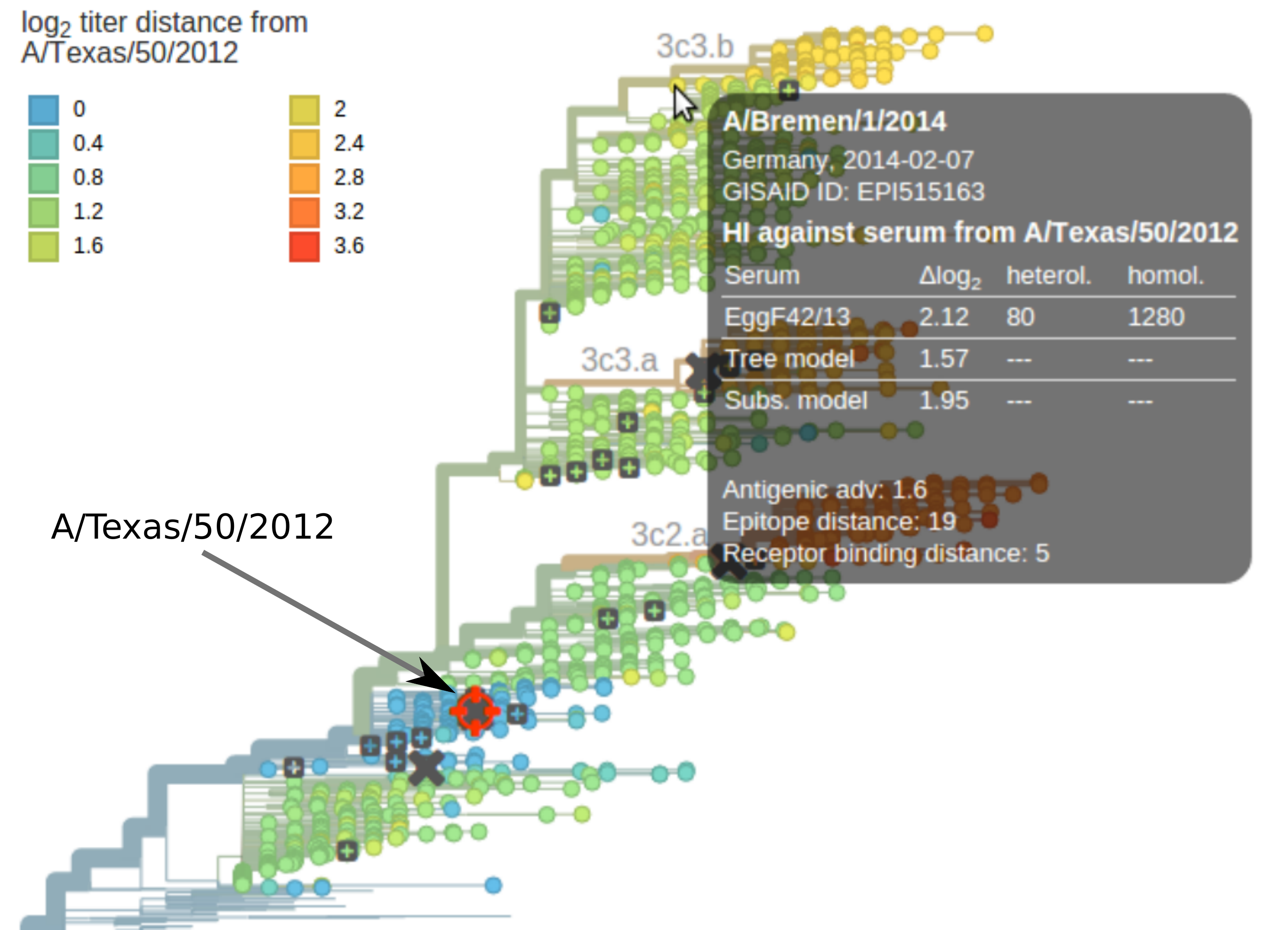}
  \caption{{\bf Visualization of antigenic evolution on the tree}. Tree tips are colored by the predicted $\log_2$ titer distance relative the focal virus (A/Texas/50/2012 in this case). The tool tip shows titers relative to all available antisera and titer predictions as a table. Each reference virus (grey squares) can be chosen by clicking with the mouse, upon which the tree color is updated. Crosses indicate vaccine strains. The tool tip shows all available measurements and the predictions by both models, while the color averages measurements relative to different antisera. 
  }
  \label{fig:titer_distance}
\end{figure}

\subsection*{Visualization of antigenic evolution}
Antigenic evolution can be visualized by mapping titer distances into the plane using a variant of multi-dimensional scaling \citep{smith_mapping_2004}. This two dimensional representation, however, is not readily superimposed with the sequence evolution of the viruses. Instead of squeezing the sequence evolution onto the plane \citep{bedford_integrating_2014}, we map the HI titer data onto the phylogenetic tree using nextflu \citep{neher_nextflu:_2015}.

The application nextflu tracks in near real time the evolution of seasonal influenza viruses and allows users to explore recent changes at particular positions, spot rapidly growing clades, and analyze the geographic distribution of viruses. We have integrated the tree and sequence models of titer data into nextflu's Python based processing pipeline augur. The titer data and the models are exported along with the tree and visualized using auspice, the JavaScript based front-end of nextflu. The resulting visualization is available at \url{HI.nextflu.org} and \FIG{titer_distance} shows a screen shot. 

This web visualization allows exploration of HI distance relative to specific antisera. All reference viruses against which antisera have been raised are indicated by grey squares. A \emph{focal reference virus} can be chosen by clicking on one of these squares, thereby coloring the tree by the average antigenic distance between viruses and antisera raised against the focal virus. Tooltips -- information boxes that pop up when the mouse hovers over a virus -- display all available measurements along with the predicted titers relative to the focal virus. The normalized and log scaled titers can be optionally corrected using estimates of antiserum potency and virus avidity. 

The tree can be colored either by measured titers (which are available only for a subset of the viruses) or the predictions by either the tree model or the substitution model. Toggling between coloring by measured and predicted titers gives an intuitive visual impression of the noise in titer data and possible inaccuracies of the model predictions. The model predictions are essentially de-noised antigenic distances that integrate information from a large number of titer measurements.

In addition to titer measurements and titer predictions, the \url{HI.nextflu.org} website allows coloring of the tree by the antigenic evolution that accumulated along branches starting from the root of the tree. The latter is similar to dimension one in antigenic cartography.

\section*{Discussion}
We have shown that antigenic evolution of seasonal influenza viruses can be accurately predicted from HA sequences using models parameterized by branches separating test and reference virus in a phylogenetic tree or by amino acid substitutions separating the two sequences. Both of these models predict titers with similar or better accuracy than more traditional cartographic approaches. The appeal of antigenic cartography \citep{smith_mapping_2004} is partly the ease with which complex tables of HI titer data can summarized, visualized, and printed on paper. To achieve similarly intuitive display of the results of the tree and substitution models, we implemented a web-browser based interactive visualization available at \url{HI.nextflu.org} \citep{neher_nextflu:_2015}. The web application integrates antigenic relationships with molecular evolution, geographic distribution, and predictors of clade growth. Our goal, rather than simply predicting antigenic data from genetic sequences, was a characterization of antigenic space, a full integration with the phylogenetic tree, and an assessment of the predictive value of HI measurements. In contrast, \citet{harvey_identifying_2014} focus on identifying specific residues in H1N1 that are responsible for titer drops, while \citep{sun_using_2013} focus on predicting titers of unknown viruses visualized via cartography.

Previous analysis had concluded that two dimensions provide the optimal embedding for antigenic evolution and adding additional dimensions did not improve the predictive power \citep{smith_mapping_2004, bedford_integrating_2014} Here we find that a model based on the tree structure -- effectively infinite dimensional -- predicts titers similar or better accuracy. This apparent discrepancy has its roots in the number of parameters necessary to specify the model. In $d$ dimensional cartography, the location of each virus is described by $d$ coordinates. Hence the number of parameters increases rapidly with the number of dimensions of the model and predictive power decreases due to overfitting. However, if distances are additive and tree-like, all $n(n-1)/2$ pairwise distances between $n$ viruses are defined by the contributions of at most $n-2$ internal and $n$ external branches (external branches are not considered by the model). In practice, the number of internal branches is substantially lower than $n-2$ due to the many polytomies in the tree. For A(H3N2) from the past 12 years, HI data was available for 1796 out of total of 2473 viruses in the tree. Of the 622 internal branches constrained by HI data, 189 branches were inferred to affect HI titers (>0.001).  
Hence with substantially fewer parameters, we achieve a better or comparable fit to the data than $d=2$ cartography, which suggests that the phylogenetic tree is a more natural space for antigenic change. We corroborated this interpretation by explicitly testing ``tree-ness'' using quartet distances and symmetry between reciprocal measurements (\FIG{symmetry}). 


We find substantial differences in the overall rate of antigenic evolution across viruses with fast antigenic drift in H3N2 and slow antigenic drift within both influenza B lineages, in agreement with \citet{bedford_integrating_2014}. Overall, antigenic drift estimated by our method tends to be somewhat lower, in particular for influenza B. This discrepancy in drift is likely due to the different model space: the phylogenetic tree provides more freedom for different clades to evolve in different directions and antigenic distances are accommodated on side branches, rather than the trunk. On a two dimensional antigenic map, however, the space for side branches and subclusters is limited, such that more antigenic distance is picked up by the backbone of the map.

Current efforts to predict the evolution and dynamics of seasonal influenza viruses \citep{luksza_predictive_2014,neher_predicting_2014,steinbruck_allele_2011,he_low-dimensional_2010,steinbruck_computational_2014} are based solely on virus HA sequences. By mapping the phenotypic HI data onto sequences and phylogenetic trees, it should be possible to improve prediction accuracy. Using HI data to predict is, however, not as straightforward as it might seem and by itself it does not predict better than the sequence-based LBI predictor \citep{neher_predicting_2014}. HA substitutions associated with large antigenic changes have a higher probability of fixation, but many causing substantial antigenic change (e.g. K158R) fail to spread in the population. Similarly, in many years clades that are antigenically more distant from previously circulating viruses die out.
The high frequency of such ``false positives'' interferes with the use of HI measurements for early detection of emerging strains that are the likely founders of future generations of the virus. We observe substantial false positives at the 1\% clade frequency level, but fewer false positives at the 5\% frequency level (\FIG{spread}B). However, using a 5\% frequency threshold loses much of the early detection capacity, limiting HI-based prediction to the regime accessible to the sequence-based genealogy approaches, such as LBI. 

The main challenge is hence to find a way to reduce the false positive rate in HI-based prediction. Success of strains with smaller HA antigenic advancement over the ones with a larger advancement could be rationalized in two ways: i) by some alternative improvement of infectivity, immune system avoidance or lower mutational load; ii) by a fitness cost associated with a large antigenic effect substitutions. Supporting the second scenario, it is known that adaptive mutations are sometimes not tolerated in certain genetic backgrounds because they destabilize the encoded protein and further, stabilizing, mutations are required to compensate for the loss of virus fitness \citep{thyagarajan_inherent_2014,gong_stability-mediated_2013}. Better understanding of the context dependence of large antigenic effect HA substitutions may therefore be a promising way towards reducing the false positive rate and improving prediction capacity. The problem of false positive detection is also seen at the Koel 7 positions. While most past dramatic changes in antigenic phenotype are associated with substitutions at these sites, substitutions at Koel 7 sites do not always have antigenic effects or fail to spread. Koel 7 substitutions alone are poor predictors.

In conclusion: our study demonstrates that HI data integrates naturally onto the sequence derived phylogeny of the virus. While at present HI-based prediction does not outperform sequence based methods, better understanding of genetic context dependence of HI data may provide a way towards improved performance. Characterizing HA substitutions that have historically been associated with antigenic transitions and placing HI data directly into genealogical context may also help with optimizing targeted acquisition of HI data.

\section*{Materials and Methods}
\subsection*{Data}
HA sequences of influenza A and B viruses isolated from humans were downloaded from GISAID. Accession numbers of all sequences are provided as supplementary files.
We collected HI data from publications \citet{smith_mapping_2004, russell_global_2008, barr_epidemiological_2010} and annual and interim reports of the WHO CC London between 2002 and 2015 \cite{NIMR02,NIMRMarch08,NIMRFeb09,NIMRFeb10,NIMRSep10,NIMRSep11,NIMRFeb12,McCauley_sep_2012,McCauley_feb_2013,McCauley_sep_2013,McCauley_feb_2014,McCauley_sep_2014,McCauley_feb_2015}. HI data prior to 2011 was curated in \citet{bedford_integrating_2014}). 
Original HI data tables are available from the website of the Worldwide Influenza Centre at the Francis Crick Institute at \url{http://www.crick.ac.uk/research/worldwide-influenza-centre/annual-and-interim-reports/}. 

While the modalities of HI assays have changed over the years (red blood cells from different species, addition of neuraminidase inhibitors, etc), we find that the model describes data spanning many years with reasonable accuracy. This insensitivity of the model is likely due to the fact that differences in HI assay methodologies can largely be absorbed in the model terms for antiserum potency and virus avidity. However the most recent data, largely provided by WHO CC London, is modeled with greater accuracy.

\subsection*{Data processing and model fit}
The data processing pipeline is based on {\bf nextflu} \citep{neher_nextflu:_2015}, which subsamples viruses, aligns sequences and builds a phylogenetic tree. This pipeline was modified for the current purpose to enforce the inclusion of all strains for which antigenic data was available. Then, in addition to the standard {\bf augur} pipeline of nextflu, the tree and substitution models were fitted to the HI data as follows.

For each combination of virus $i$ and antiserum $\alpha$, we define antigenic distance as $H_{i,\alpha} = \log_2(T_{a\alpha}) - \log_2(T_{i\alpha})$, where $T_{i\alpha}$ is the antiserum titer required to inhibit virus $i$ and $T_{a\alpha}$ is the homologous titer. 
In case multiple measurements are available, we average the base 2 logarithm of the titers. When no homologous titer was available, the maximal titer was used as a proxy for the homologous titer.
The path between virus $i$ and antiserum $\alpha$ extends over branches of the phylogeny $b \in (i \ldots \alpha)$, where each branch has a branch effect $d_b$.
The tree distance $\Delta_{i, \alpha}$ between virus $i$ and antiserum $\alpha$ is defined in \EQ{model}.
The parameters $d_b$, $p_\alpha$, $c_i$ are then estimated by minimizing the cost function
\begin{equation}    
\label{eq:cost}
C = \sum_{i,\alpha} (H_{i,\alpha} - \Delta_{i,\alpha})^2 + \lambda \sum_b d_b + \gamma \sum_i c_i^2 + \delta \sum_\alpha p_\alpha^2
\end{equation}
subject to the constraints $d_b\geq0$. To avoid overfitting, the different parameters of the model are regularized by the last three terms in \EQ{cost}. Large titer drops are penalized with their absolute value multiplied by $\lambda$ ($\ell_1$ regularization), which results in a sparse model in which most branches have no titer drop \citep{candes_decoding_2005}. Similarily, the antiserum and virus avidities are $\ell_2$-regularized by $\gamma$ and $\delta$, penalizing very large values without enforcing sparsity. This constrained minimization can be cast into a canonical convex optimization problem and solved efficiently, see below. In the substitution model, the sum over the path in the tree is replaced by a sum over amino acid differences in HA1. Sets of substitutions that always occur together are merged and treated as one compound substitution. The inference of the substitution model parameters is done in the same way as for the tree model (see \citet{harvey_identifying_2014,sun_using_2013} for a similar approach). Since there are only a small number of antisera and differences in antiserum potency are often on the order of one or two antigenic units, $\delta$ was assigned a small value of 0.3, while $\lambda=1.0$ was used to regularize branch or substitution and $\gamma=2.0$ for virus effects. The quality of the fit depends weakly on these parameters.

The total number of adjustable parameters are $S$ antiserum potencies, $V$ avidities of viruses, and $M$ internal branches of the tree of which there are at most $V-1$, but typically fewer due to many polytomies in the trees. In practice only a fraction of the branches have non-zero branch effects and the total number of non-zero parameters is not much larger than the number of test viruses. In the substitution model, the number of non-singleton substitutions found in an HA1 alignment is typically on the order of 100, most of which are inferred to have no antigenic effect.

This optimization of \EQ{cost} can be cast into a canonical quadratic programming problem of the form
\begin{eqnarray}
\label{eq:QP}
\mathrm{minimize:} &\quad \frac{1}{2}\mathbf{xQx} + \mathbf{qx}  &\\
\mathrm{subject\,to:}& \mathbf{Gx}\leq\mathbf{h} \nonumber
\end{eqnarray}
where $\mathbf{x}$ is the vector of unknowns, and the matrix $\mathbf{Q}$ and the vector $\mathbf{q}$ specify the cost function. The matrix $\mathbf{G}$ and the vector $\mathbf{h}$ encode inequality constraints on $\mathbf{x}$.

To formulate \EQ{cost} in this canonical form, we concatenate the titer drops $d_b$ associated with branches $b$ of the tree, the potencies $p_\alpha$ of each antiserum, and avidities $c_i$ of virus isolates into a single vector $\mathbf{x}$
\begin{equation}
  x_s = \begin{cases}
    d_s, &\text{ for } s=1\ldots B\\
    c_s, &\text{ for } s=B+1\ldots B+V\\    
    p_s, &\text{ for } s=B+V+1\ldots B+V+S\\    
  \end{cases}
\end{equation}
where $B$ are the number of branches in the tree connecting measurements, $V$ is the number of viruses with titer measurements, and $S$ are the number of antisera. Next, we construct a large binary matrix $\A$ of dimension $N\times(B+V+S)$, where $N$ is the total number of measurements. Each row of the matrix $\A$ codes for a titer prediction $\Delta_{i\alpha}$. Using the double index $i\alpha$ to label measurements, entries of $\A$ are given by 
\begin{equation}
  \A_{i\alpha, s}=\begin{cases}
    1, &\text{ for }s\text{ in path, potency, or avidity}\\
    0, &\text{ otherwise } 
  \end{cases}
\end{equation}
i.e., $\A_{i\alpha,s}=1$ for all $s$ that correspond a branch of the path $(i\ldots\alpha)$, the virus avidity $c_i$ and the antiserum potency $p_\alpha$; otherwise, $\A_{i\alpha,s}=0$. All titer predictions of \EQ{model} are hence given by  
\begin{equation}
  \mathbf{\Delta} = \A \mathbf{x}
\end{equation}
where $\mathbf{x}$ is the vector of parameters. 

Using these definitions, the cost function \EQ{cost} can be written as 
\begin{equation}
(\mathbf{H} - \A\mathbf{x})^{T}(\mathbf{H} - \A\mathbf{x}) + 2\lambda \sum_{s=1}^B x_s + \gamma \sum_{s=B+1}^{B+V} x_s^2 + \delta \sum_{s=B+V+1}^{B+V+S} x_s^2
\end{equation}
subject to $x_s\geq 0$ for $s=1 \ldots B$. Dropping constant terms and defining 
\begin{equation}
  Q_{rs} = \begin{cases}
    \sum_i A_{ir}A_is, &\text{ if }r,s\leq B\\
    \gamma &\text{ if }r=s, s>B, s\leq B+V\\
    \delta &\text{ if }r=s, s>B+V\\
    0&\text{ otherwise}
  \end{cases}
\end{equation}
we have 
\begin{equation}
  \mathbf{x}\mathbf{Q}\mathbf{x} - 2\mathbf{H}\A\mathbf{x} + 2\lambda \sum_{s=1}^B x_s
\end{equation}
To enfore the $\ell_1$ regularization and the positivity of the titer drops corresponding to $x_s$, $s=1\ldots B$, we define inequality constraints
\begin{equation}
  \label{eq:inequality}
  \begin{split}
    x_s\geq & 0 \\
  \end{split}
\end{equation}
for $s=1\dots B$ which forces all titer drops to be positive. In addition, we set 
\begin{equation}
  \mathbf{q} = -\mathbf{H}\A\mathbf{x}
\end{equation}
and add $\lambda$ to $q_{s},\,s=1,\dots B$ to penalize large effects. With these definitions, we have cast \EQ{cost} in the form of \EQ{QP}. The resulting quadratic programming problem is then solved with \texttt{cvxopt} by M.~Andersen and L.~Vandenberghe.

The HI titer data and the inferred model parameters are integrated into the json data structure describing the tree or saved in an additional data file for later visualization using {\bf auspice}.

\subsection*{Visualization}
The HI titer coloring and tool tips is implemented via straightforward extension of nextflu/auspice functionality. In addition to the standard nextflu tree display, a structure showing the positions at which the substitution model inferred large contribution to antigenic change are shown on the pages for each individual virus lineage. The structures are visualized with JSmol \citep{jsmol}. For H3N2, we use structure 5HMG \citep{weis_refinement_1990}, for H1N1pdm we use 4LXV \citep{yang_structural_2014}, for the influenza B Victoria and Yamagata lineages we use  4FQM \citep{dreyfus_highly_2012} and 4M40 \citep{ni_structural_2013}.

\section*{Acknowledgements}
We gratefully acknowledge the network of WHO National Influenza Centres, comprising the WHO Global Influenza Surveillance and Response System, for providing the influenza viruses used in this study and the WHO CCs that produced the wealth of HI titer data analyzed here. 
We are grateful to John McCauley for making HI titer data available and for valuable feedback on this manuscript.
The work of the WIC (RSD) was supported by the Medical Research Council under programme number U117512723. We also acknowledge the authors, originating and submitting laboratories of the sequence data, downloaded from GISAID’s EpiFlu Database, on which this research is based. A full list of all laboratories who contributed to the data used here is available at \url{HI.nextflu.org/acknowledgements}. This work is supported by the ERC though Stg-260686, by the NIH through U54 GM111274, by the Simons Foundation Grant \#326844, and by a University Research Fellowship from the Royal Society.

\bibliography{titer}

\appendix
\setcounter{figure}{0}
\setcounter{table}{0}
\renewcommand{\thetable}{S\arabic{table}}
\renewcommand{\thefigure}{S\arabic{figure}}

\begin{table}[tb!]
\scriptsize
\centering  
\begin{tabular}[b]{|l|c|c|c|c|c|}
  \hline
  \hline
\textbf{substitutions}&	\textbf{85 to 95} & \textbf{90 to 00} & \textbf{95 to 05} & \textbf{00 to 10} & \textbf{05 to 16}\\ \hline
K189N & --- & --- & --- & 3.44 & 2.65\\ 
K158N/N189K & --- & --- & --- & 2.17 & 3.25\\ 
K156E & --- & 3.18 & --- & --- & ---\\ 
C1$^{*}$ & --- & 3.03 & --- & --- & ---\\ 
S262N & --- & 2.61 & 0.0 & --- & 0.04\\ 
K135G & 1.9 & 2.17 & --- & --- & ---\\ 
C1R$^{*}$ & --- & 2.1 & --- & --- & ---\\ 
K158R & --- & --- & --- & 1.91 & 2.08\\ 
C2$^{*}$ & --- & --- & 1.94 & --- & ---\\ 
N145K & 1.88 & 0.01 & 0.42 & 0.5 & ---\\ 
K193N & 1.67 & --- & --- & --- & ---\\ 
K140E & --- & --- & 1.65 & 1.41 & ---\\ 
H155Y/R189K & 1.57 & --- & --- & --- & ---\\ 
T135G & --- & 1.55 & --- & --- & ---\\ 
K145N & 0.47 & 0.0 & 1.53 & 1.18 & ---\\ 
S133D/E156K & --- & 1.49 & --- & --- & ---\\ 
S157L & 0.47 & 0.79 & --- & --- & 1.4\\ 
Y155H/K189R & 1.4 & --- & --- & --- & ---\\ 
K135T & --- & 1.37 & --- & --- & ---\\ 
S193F/D225N & --- & --- & --- & 1.35 & ---\\ 
T212A & --- & --- & --- & 1.27 & 0.62\\ 
C3$^{*}$ & 1.24 & --- & --- & --- & ---\\ 
N144K & --- & --- & --- & 1.24 & 0.0\\ 
K144D & --- & --- & --- & --- & 1.23\\ 
N121T & --- & --- & 1.21 & --- & ---\\ 
K62E/N276K & --- & --- & 1.19 & --- & ---\\ 
S159Y & 1.17 & --- & --- & --- & 0.83\\ 
S133D & 1.17 & --- & --- & --- & ---\\ 
S193N & 1.16 & --- & 0.56 & 0.0 & ---\\ 
L226V & --- & 0.19 & 1.13 & --- & ---\\ 
S189R & --- & 1.1 & --- & --- & ---\\ 
G186S & --- & --- & 1.07 & --- & ---\\ 
E190D & 0.0 & 1.06 & --- & --- & ---\\ 
K135E & 1.04 & 1.05 & --- & --- & ---\\ 
Q156H & --- & --- & 1.04 & 0.67 & ---\\ 
K144N & --- & --- & --- & --- & 1.03\\ 
T135K & --- & 0.0 & 1.01 & --- & ---\\ 
Y159F & --- & --- & 1.01 & 0.9 & 0.0\\ 
I112V/S193F & --- & --- & --- & --- & 1.01\\ 
T212S & --- & --- & --- & --- & 1.0\\ 
K140I & --- & --- & --- & 0.74 & 1.0\\ 
K135D & --- & 0.99 & --- & --- & ---\\ 
K156E/D190E & 0.97 & --- & --- & --- & ---\\ 
F159Y & --- & --- & 0.17 & 0.34 & 0.97\\ 
C4$^{*}$ & --- & --- & 0.97 & 0.27 & ---\\ 
D172G/Q197R/N278S & 0.96 & --- & --- & --- & ---\\ 
T214I & 0.11 & 0.91 & --- & --- & ---\\ 
R189S & 0.9 & 0.7 & --- & --- & ---\\ 
K156H & --- & --- & 0.9 & --- & ---\\ 
L157S/S189R & 0.89 & --- & --- & --- & ---\\ 
  \hline
  \hline
  \end{tabular}
  \caption{The 50 largest inferred antigenic effects of substitutions in HA1 for A/H3N2 in the past 30 years as inferred by the substitution model in overlapping 10 year intervals. When several substitutions always occurred together, a combined effect is shown. A dash indicated the absence of the substitution in a particular time interval. The substitutions are sorted by the maximum across time intervals.
$^{*}$ C1: \texttt{E62K, N121T, S124G, N133D, R142G, I144V, Q156K, K158E, A196V, K276N}; C1R: \texttt{K62E, V144I, K156Q, E158K, V196A, N276K}, i.e., largely the reverse of cluster C1; C2: \texttt{S124G, N133D, I144V, Q156K, K276N} a subset of cluster C1; C3: \texttt{K82E, E83K, A131T, R299K}; C4: \texttt{I25L, Q75H, T131A, T155H}. }.
  \label{tab:aminoacids}
  \label{fig:structure}
\end{table}

\begin{figure*}[tb]
  \centering
  \includegraphics[width=1.9\columnwidth]{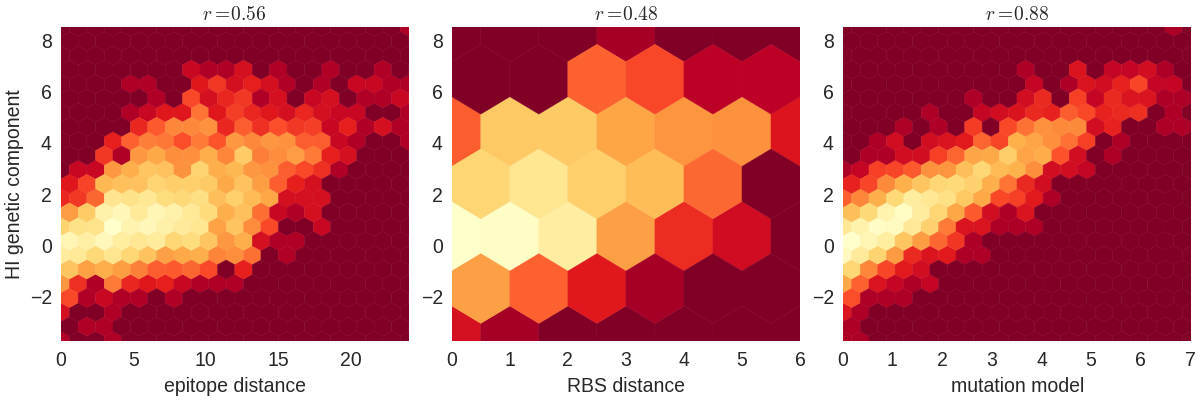}
  \caption{Correlation of HI titer measurements (y-axis) with epitope distance between test and reference virus, distance at receptor binding sites, and the genetic component of the substitution model. Titer measurements are corrected by antiserum potency and virus avidity. Neither amino acid, epitope, or RBS distance explain much of the titer variation. }
  \label{fig:distance_correlations}
\end{figure*}

\begin{figure*}[tb]
  \centering
  \includegraphics[width=1.\columnwidth]{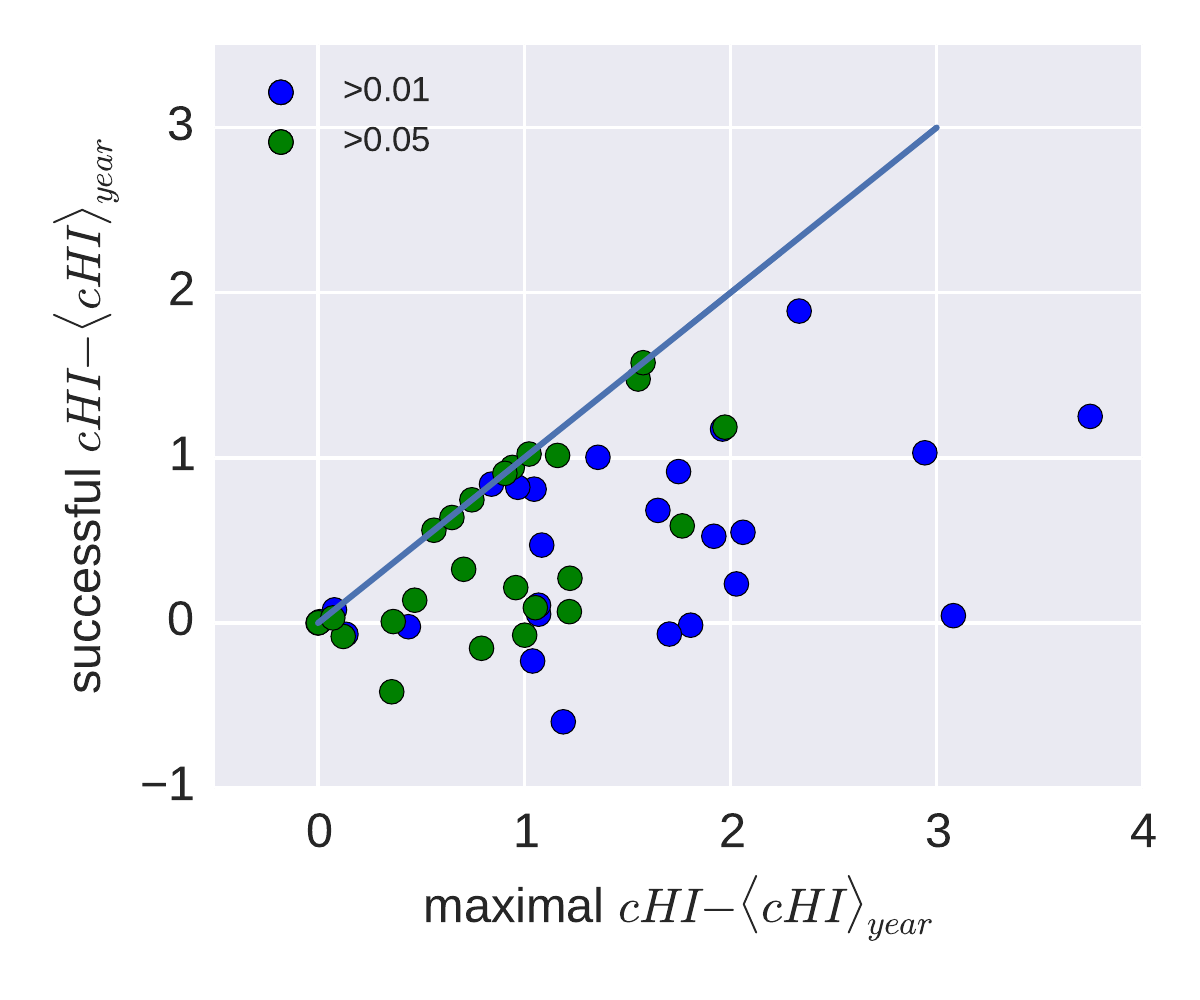}
  \caption{The in many years, the most antigenically advanced clade is not the clade dominating the next season. The figure shows the maximally observed centered cHI vs the centered cHI of the successful clade for each year from 1990 to 2014. }
  \label{fig:best_HI_vs_HI_of_best}
\end{figure*}

\begin{figure*}[tb]
  \centering
  \includegraphics[width=1.\columnwidth]{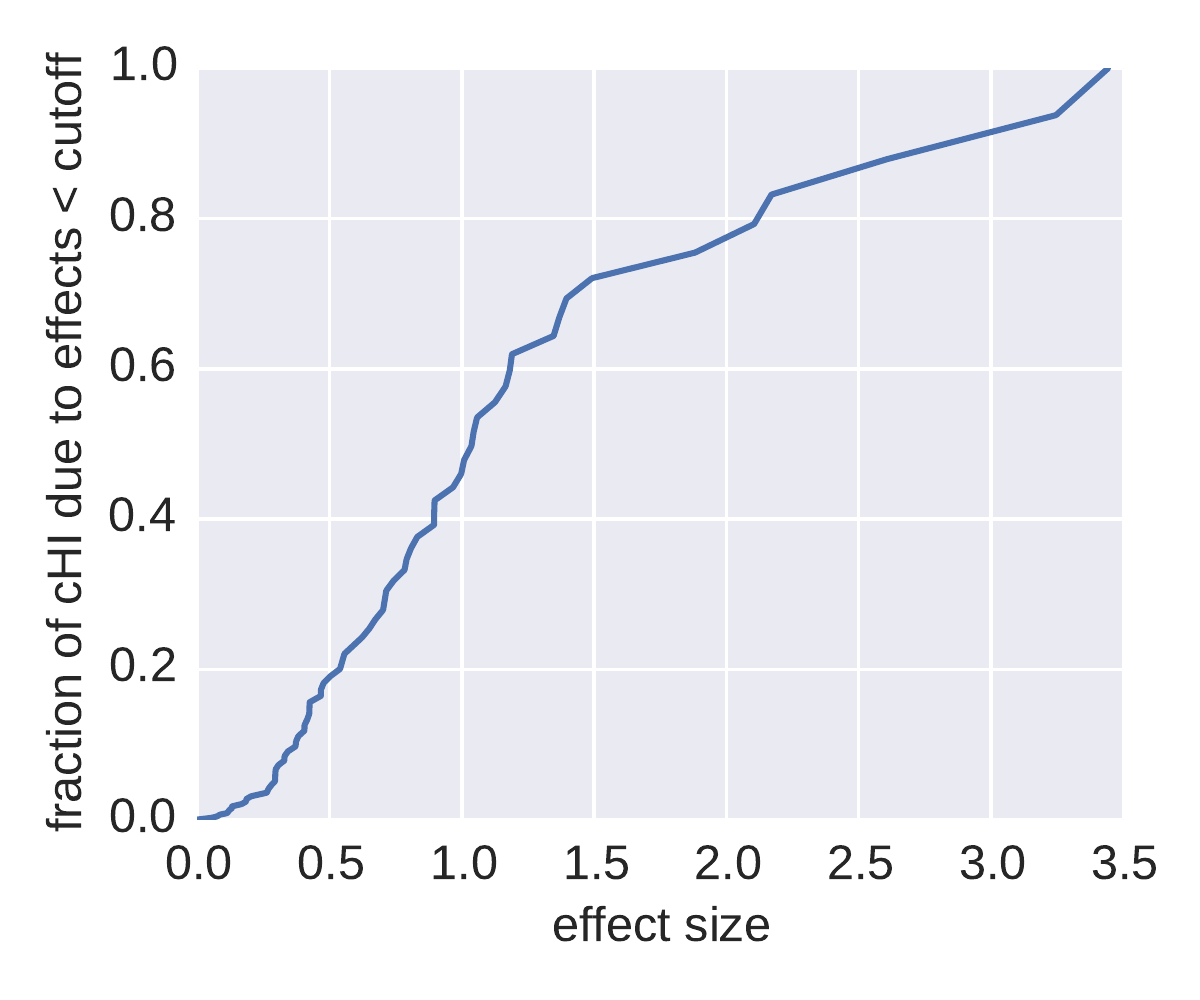}
  \caption{About half of antigenic change is attributed to mutations with effects smaller than one unit. The figure shows the fraction antigenic change by common mutations with effects smaller than the cutoff on the horizontal axis.}
  \label{fig:cumulative_effects}
\end{figure*}

\end{document}